\documentclass[twocolumn]{aastex62}

\usepackage{longtable, graphicx, subfigure, comment}
\usepackage[]{hyperref}
\usepackage[fleqn]{amsmath}
\usepackage{lmodern}
\usepackage[T1]{fontenc}
\usepackage{url}
\usepackage{todonotes}

\newcounter{column_number}
\newcommand{\MSun}{\ifmmode {M_{\odot}}\else{M$_{\odot}$}\fi}
\newcommand{\lapprox }{{\lower0.8ex\hbox{$\buildrel <\over\sim$}}}
\newcommand{\gapprox }{{\lower0.8ex\hbox{$\buildrel >\over\sim$}}}

\newcommand\kms{\ensuremath{\mbox{km}\,\mbox{s}^{-1}}}
\newcommand\Teff{\ensuremath{T_\mathrm{eff}}}
\newcommand\logg{\ensuremath{\log g}}
\newcommand\vsini{\ensuremath{v\sin i}}

\newcommand\met{[M/H]}
\newcommand\feh{[Fe/H]}
\newcommand\vmic{$\xi_t$}

\received{, 2018}
\revised{, 2018}
\accepted{, 2018}
\submitjournal{ApJ}

\shorttitle{Automated Abundances for Five Praesepe GK Stars}
\shortauthors{Gebran et al.}

\begin{document}

\title{Pushing Automated Abundance Derivations Into the Cool Dwarf Regime:\\A Test Using Three G and Two K Stars in Praesepe\footnote{Based on observations obtained with the Apache Point Observatory 3.5-meter telescope, which is owned and operated by the Astrophysical Research Consortium.}}

\correspondingauthor{Marwan Gebran}
\email{mgebran@ndu.edu.lb}

\author[0000-0002-8675-4000]{Marwan Gebran}
\affiliation{Department of Physics and Astronomy, Notre Dame University-Louaize, PO Box 72, Zouk Mika\"el, Lebanon}
\affiliation{Department of Astronomy, Columbia University, 550 West 120th Street, New York, NY 10027, USA}

\author[0000-0001-7077-3664]{Marcel A.~Ag{\"u}eros}
\affiliation{Department of Astronomy, Columbia University, 550 West 120th Street, New York, NY 10027, USA}

\author[0000-0002-1423-2174]{Keith Hawkins}
\affiliation{Department of Astronomy, The University of Texas at Austin, 2515 Speedway Boulevard, Austin, TX 78712, USA}

\author[0000-0001-7203-8014]{Simon C.~Schuler}
\affiliation{University of Tampa, Department of Chemistry, Biochemistry, and Physics, Tampa, FL 33606, USA}

\author[0000-0003-2528-3409]{Brett M.~Morris}
\affiliation{Department of Astronomy, University of Washington, Box 351580, Seattle, WA 98195, USA}



\begin{abstract}
We present the results of an abundance analysis of three G and two K dwarfs in the Praesepe open cluster based on high-resolution, moderate signal-to-noise-ratio spectra obtained with the ARC 3.5-m Telescope at Apache Point Observatory. Using a Principle Component Analysis and the BACCHUS automated spectral analysis code, we determined stellar parameters and abundances of up to 24 elements for each of our targets, which range in temperature from 6000 to 4600 K. The average derived iron abundance for the three G stars is 0.17$\pm$0.07~dex, consistent with the 0.12$\pm$0.04~dex derived by \cite{boesgaard2013} for their sample of 11 solar-type Praesepe members, which included these G stars. To investigate the efficacy of using automated routines to derive the abundances of cooler main-sequence stars, we compared the abundances of the K dwarfs to those of the G dwarfs. Our abundances are consistent to $\leq$0.1~dex for 13 of the 18 elements we report for all five of the stars, providing more evidence that G and K stars in a given open cluster are chemically homogeneous. The median difference between the mean G and K stars abundances is 0.08$\pm$0.05 dex, despite serious challenges with the noisier data for the fainter K dwarfs. Our results are encouraging for chemical tagging, as they indicate that it may be possible to use automated abundance determination techniques to identify chemically related main-sequence stars across larger temperature ranges than are usually considered in these experiments.
\end{abstract}


\section{Introduction} 
\label{sec:intro}
The recent advent of large-scale stellar spectroscopic surveys dedicated to the systematic determinations of abundances of many elements was in part inspired by the concept of chemical tagging, a process by which distinct stellar populations in the Galaxy might be identified based on elemental abundances \citep{2002ARA26A..40..487F,2006AJ....131..455D,blandhawthorn2010}. This idea originates in the notion that stars born together should share a chemical fingerprint that is sufficiently unique to distinguish these stars, even if no longer obviously associated, from the Galactic background.

\begin{table*}[!th]
\begin{center}
\caption{APO observations}\label{log_obs}
\begin{tabular}{|c|c|c|c|c|}\hline
Star & RA, Dec & $V$   & Date & Exposure time\\
     & (J2000) & (mag) &      & (s) \\ 
\hline
JS 482 & 08 42 20.09 $+$19 09 05.67 & 13.52 & 11/28/2017 & 7200 \\
JS 552 & 08 43 56.72 $+$19 43 32.29 & 13.26 & 12/29/2017 & 8400 \\
KW23  & 08 37 11.49 $+$19 48 13.25 & 11.29 & 03/30/2018 & 1800\\
KW30  & 08 37 22.23 $+$20 10 37.24 & 11.40 & 03/30/2018 & 1800 \\ 
KW208 & 08 39 45.75 $+$19 22 01.15 & 10.66 & 03/30/2018 & 1800\\ \hline 
\end{tabular}
\end{center}
\end{table*}

There are a number of reasons to think that chemical tagging should work. Studies of solar-type wide binary systems and of open clusters, which are thought to be co-eval and co-chemical populations, have shown that their component stars have very similar abundances \citep[e.g.,][]{Andrews2018, Bovy2016}, with differences in individual elements likely due to measurement uncertainties. Moreover, tests designed to identify stellar populations based on their chemical similarities have shown that they can recover stars that are associated in phase space or that appear to have a common origin \citep[e.g.,][]{Hogg2016, 2017MNRAS.465..501S}. 

Equally, however, there are reasons to be skeptical that chemical tagging can work. Open clusters, for example, may have chemical-abundance patterns that are difficult to distinguish from each others' \citep{2015A&A...577A..47B}, or indeed that are insufficiently distinct from those of unassociated field stars \citep[so-called doppelg\"angers;][]{Ness2018}. Dynamical processes such as planet formation may also alter the photospheric compositions of stars during their main-sequence lifetimes, potentially resulting in chemical inhomogeneities within what was once a chemically homogeneous group \citep[e.g.,][]{2018ApJ...863..179S}. Such signatures are generally detectable only with very high precision, typically $\leq$0.02~dex \citep[e.g.,][]{2016MNRAS.457.3934L,2016AJ....152..167T}, however, and may not be important for large-scale surveys.


Another potential concern for chemical tagging as applied to main-sequence stars is the derived abundance anomalies for some elements in open cluster late-G and K dwarfs. The abundances of iron derived from singly-ionized lines and oxygen derived from the near-infrared triplet, in particular, demonstrate a striking increase with decreasing temperature for stars with $T_{\mathrm{eff}} \lesssim 5300 \, \mathrm{K}$ in the Hyades, Pleiades, and M34 \citep[e.g.,][]{2003AJ....125.2085S,2004ApJ...602L.117S,2004ApJ...603..697Y} and in the field \citep[e.g.,][]{2007A&A...465..271R}. The source of the discrepant abundances has yet to be fully identified, but line-blending in the spectra of cooler stars has been suggested to account for some but not all of their observed overabundances \citep{2017ApJ...846...24A}. Studies have also investigated the influence of metallicity and chromospheric activity on the abundances, with no statistically significant correlations being found \citep[e.g.,][]{2006ApJ...636..432S,2017ApJ...846...24A}. \citet{2006ApJ...636..432S} reported a possible age-related diminution of the cool-star abundances. These authors also demonstrated that photospheric temperature inhomogeneities (i.e., starspots) in  cool stars are a plausible source of the overabundances. Such uncertainty in the abundances derived for late-G and K dwarfs may be an indication of deficiencies in current cool-star model atmospheres. 

To further explore these ideas and challenges, we obtained $R\approx$ 31,000 spectra (described in Section~\ref{sec:obs}) of five main-sequence members of the Praesepe open cluster. In this paper, we describe our application of some of the techniques generally applied to higher-mass/evolved stars to derive fundamental properties of, and abundances for, these three G and two K dwarfs (Sections~\ref{sec:param} and \ref{sec:abun}). We then use our results to derive a cluster metallicity and to examine the homogeneity of abundances for a range of elements in $\leq$1~\MSun\ stars in Praesepe (Section~\ref{sec:disc}). We conclude that chemical tagging remains plausible, but challenging, for main-sequence stars given our data and the techniques currently available.

\section{Observations and data reduction} \label{sec:obs}
Praesepe, the Beehive Cluster (M44; NGC2632), is a nearby \citep[$\approx$180 pc;][]{vanleeuwen2009}, rich \citep[$\approx$1200 stars;][]{adam2007}, and intermediate-age \citep[$\approx$600~Myr;][]{delorme2011} open cluster. Praesepe's proximity makes it an ideal target for observers interested in the properties of a single-aged stellar population, including multiplicity \citep[e.g.,][]{hillenbrand2018} and angular-momentum evolution \citep[e.g.,][]{douglas2017}, and in circumstellar processes such as planet formation \citep[e.g.,][]{mann2017, rizzuto2018}.

Our spectroscopic observations of five high-confidence Praesepe members\footnote{All five have membership probabilities $>$99\% in the \citet{adam2007} cluster catalog.} were obtained with the ARC Echelle Spectrograph (ARCES), an $R\approx$ 31,000 cross-dispersed spectrograph, mounted on the ARC 3.5-m telescope at the Apache Point Observatory, NM (see Table~\ref{log_obs}). 

We began by observing two K dwarfs, JS 482 (EPIC 211928486) and JS 552 (EPIC 211966629), on the nights of 2017 November 28 and December 29, respectively. Exposures of 7200 and 8400 s achieved signal-to-noise ratios (SNRs) of $\approx$50-60. The three G dwarfs, KW23, KW30, and KW208, were observed on the night of 2018 March 30. Each of these exposures was 1800 s and the corresponding SNRs $\approx$ 80-100. 

We reduced our spectra using standard \texttt{IRAF}\footnote{IRAF is distributed by the National Optical Astronomy Observatory, which is operated by the Association of Universities for Research in Astronomy (AURA) under a cooperative agreement with the National Science Foundation.}\citep{1986SPIE..627..733T} procedures that subtract biases, remove cosmic rays, normalize by the flat field, and do the wavelength calibration with exposures of a Th-Ar lamp.\footnote{An ARCES data reduction manual is available here: \url{http://astronomy.nmsu.edu:8000/apo-wiki/attachment/wiki/ARCES/Thorburn_ARCES_manual.pdf}} To measure the blaze function, we fit the spectrum of an early type star with a high-order polynomial and  divided the spectrum of each target by the polynomial fit to normalize each spectral order. To shift the normalized spectra into the rest-frame by removing the stellar radial velocities, we maximized the cross-correlation of the ARCES spectra with PHOENIX \citep{Husser2013,Hauschildt3,Hauschildt2,Hauschildt1} model spectra. All these reduction steps are implemented in the automated software of \cite{2018JOSS....3..854M}.

\section{Determining the Fundamental Stellar parameters} \label{sec:param}
Abundance analyses are very sensitive to the inputed fundamental stellar parameters, especially the effective temperature (\Teff), the surface gravity (\logg), and the microturbulent velocity ($\xi_t$). These should therefore be derived as accurately as possible. For that reason, we employed three different techniques to obtain the fundamental parameters of our target stars: a Principal Component Analysis (PCA), iron-excitation-ionization balance, and photometry. These approaches and their results are described below.

\subsection{Using Principal Component Analysis}
We started by applying the PCA technique of \cite{Gebran} to our five stars. PCA is a technique that has been rarely used in stellar spectroscopy, but \cite{Gebran} showed that it could be used to derive accurate parameters for A stars using synthetic learning databases. \cite{S4n} used PCA to successfully derive the fundamental parameters of a sample of FGK stars using a learning database composed of observed spectra with well-known and reliable parameters. 

We derived the parameters of our observed G and K dwarfs using purely synthetic data. First, a learning database of synthetic spectra was calculated using the \texttt{SYNSPEC48} code \citep{spectra} in the 5000$-$6000~\AA\ wavelength range. This region, which is free of molecular bands, was chosen because of the presence of the prominent Mg\,{\sc i} triplet and the Na\,{\sc i} doublet, which are both sensitive to \Teff\ and \logg. In addition, the presence of a large number of weak metallic lines in this region provide a good indication of \vsini\ and of the metallicity (\met).\footnote{\met\ is an indicator of the overall elemental abundance and not just that of iron (\feh).} 

The ranges for the various stellar parameters of the spectra in the learning database are displayed in Table~\ref{range-param}. We fixed the microturbulent velocity $\xi_t = 1.0$~\kms, as that is the average value for GK dwarfs \citep{micro-FGK} and the selected wavelength region is weakly affected by the variation of $\xi_t$. 

\begin{figure*}[!th]
\centerline{\includegraphics[width=2.19\columnwidth]{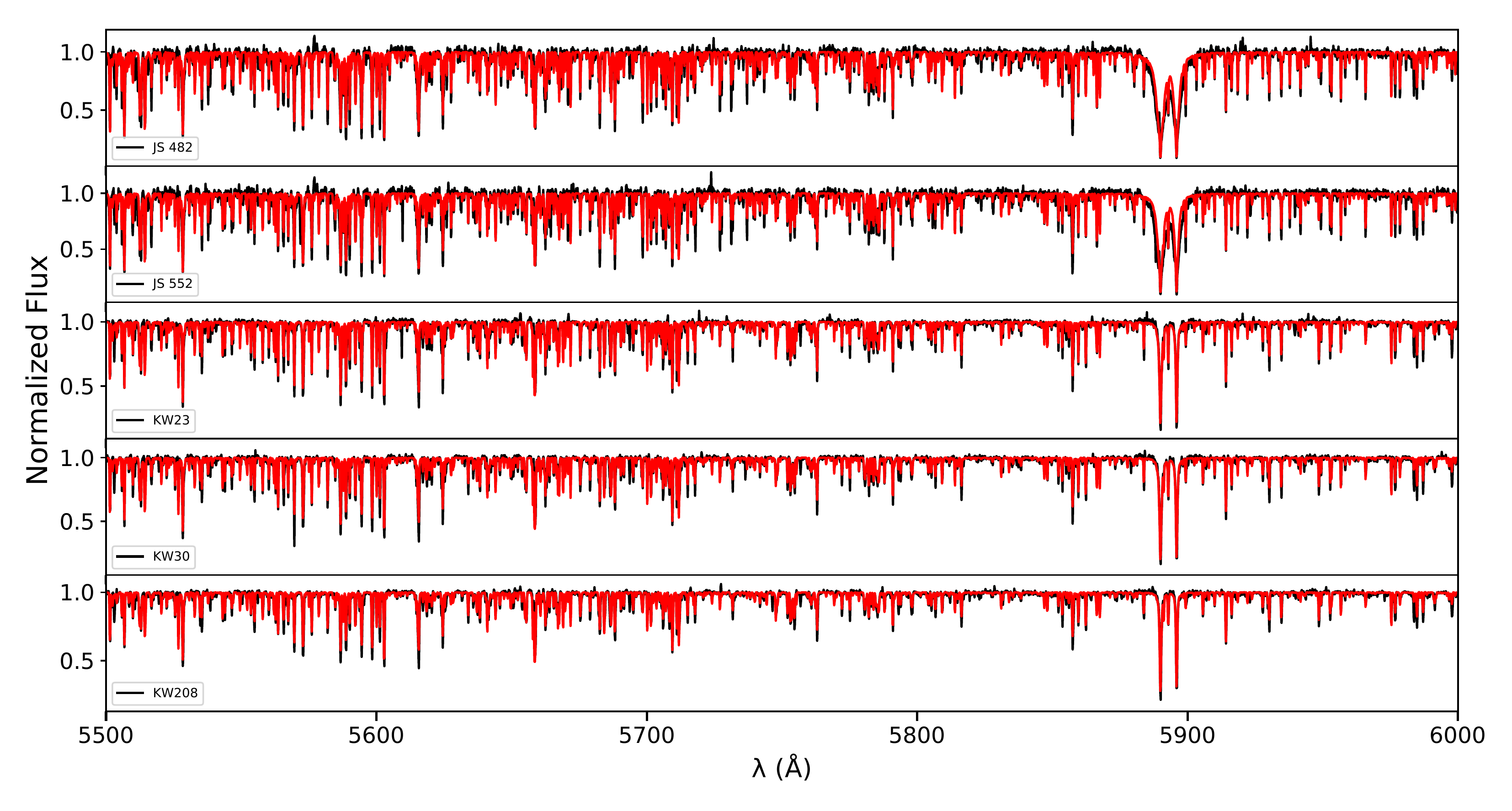}}
\caption{Fits to our spectra for the K (top two) and G (bottom three) dwarfs in the 5500$-$6000 \AA\ range. The black spectra are the data and the red spectra are the best fits calculated according to the values derived from the PCA technique.}\label{Fig-PCA}
\end{figure*}

\begin{table}
\center
\caption{Parameter ranges of the learning database used for the PCA analysis}\label{range-param}
\begin{tabular}{|c|c|c|}
\hline
& Range& Step\\ \hline
\Teff & 4000$-$6000 K & 100 K\\
\logg & 4.00$-$5.00 dex & 0.10 dex\\
\met  & -0.50$-$0.5 dex & 0.10 dex\\
$\xi_t$ & 1.0 \kms & --\\
\vsini & 0$-$15 \kms & 1 \kms\\
Resolution & 31,500 & --\\ 
$\lambda$ & 5000$-$6000 \AA & 0.15 \AA \\ \hline
\end{tabular}
\end{table}

We adopted the Kurucz gfhyperall.dat\footnote{\url{http://kurucz.harvard.edu}} line lists, which we modified with more recent and accurate atomic data retrieved from the
VALD\footnote{\url{http://www.astro.uu.se/~vald/php/vald.php}} and the
NIST\footnote{\url{http://physics.nist.gov}} databases (for more details, see \citealt{Gebran}). The structure of the atmosphere of these synthetic models were calculated using \texttt{ATLAS9} code \citep{Kurucz1992}. We used the new opacity distribution function of \cite{castelli}, calculated for $\xi_t = 1.0$~\kms. These models assume local thermodynamic equilibrium, hydrostatic equilibrium, radiative equilibrium, and a 1D plane-parallel atmosphere. Convection was treated using a mixing length parameter of 1.25 \citep{smalley2004}.

Figure~\ref{Fig-PCA} displays the observed spectra (in black) of our G and K dwarfs as well as the best-fit synthetic spectra (in red). The best fit is calculated using the stellar parameters derived from the first neighbor that minimizes the function:
\begin{equation}
d_j^{(O)}=\Sigma_{k=1}^{12}(\rho_k - p_{jk})^2,
\end{equation}
where $\rho_k$ and $p_{jk}$ are the coefficients of the projection of the flux of the observation and the $j^{th}$ synthetic spectrum, respectively, on the $k^{th}$ eigenvector of the synthetic spectra covariance matrix. Based on an error reconstruction verification, as done in \cite{Gebran}, the maximum number of eigenvector was fixed to be 12. The derived \Teff, \logg, \met\ and \vsini\ for our five Praesepe targets are given in Table~\ref{results-param}. The associated errors are upper limits on the derived parameters of the $\approx$250 FGK stars to which \cite{S4n} applied the same PCA. 

\subsection{Using Iron-Excitation-Ionization Balance}
\label{subsec:Febalance}
We also employed the Brussels Automatic Code for Characterizing High accUracy Spectra \texttt{BACCHUS} \citep{Masseron2016} to derive the fundamental parameters of our target stars. BACCHUS uses the \texttt{MARCS} \citep{Gustafsson2008} model atmosphere grid, and the radiative transfer code \texttt{TURBOSPECTRUM} \citep{Alvarez1998, Plez2012}. Atomic lines are taken from the most recent (fifth) version of the {\it Gaia}-ESO linelist (Heiter et al., in preparation). Molecular species are also included for CH \citep{Masseron2014}, and CN, NH, OH, MgH and C$_2$; the lines of SiH molecules are adopted from the Kurucz linelists and those for TiO, ZrO, FeH, CaH from B.~Plez (private communication). 

\begin{figure}[!th]
\centerline{\includegraphics[width=1.02\columnwidth]{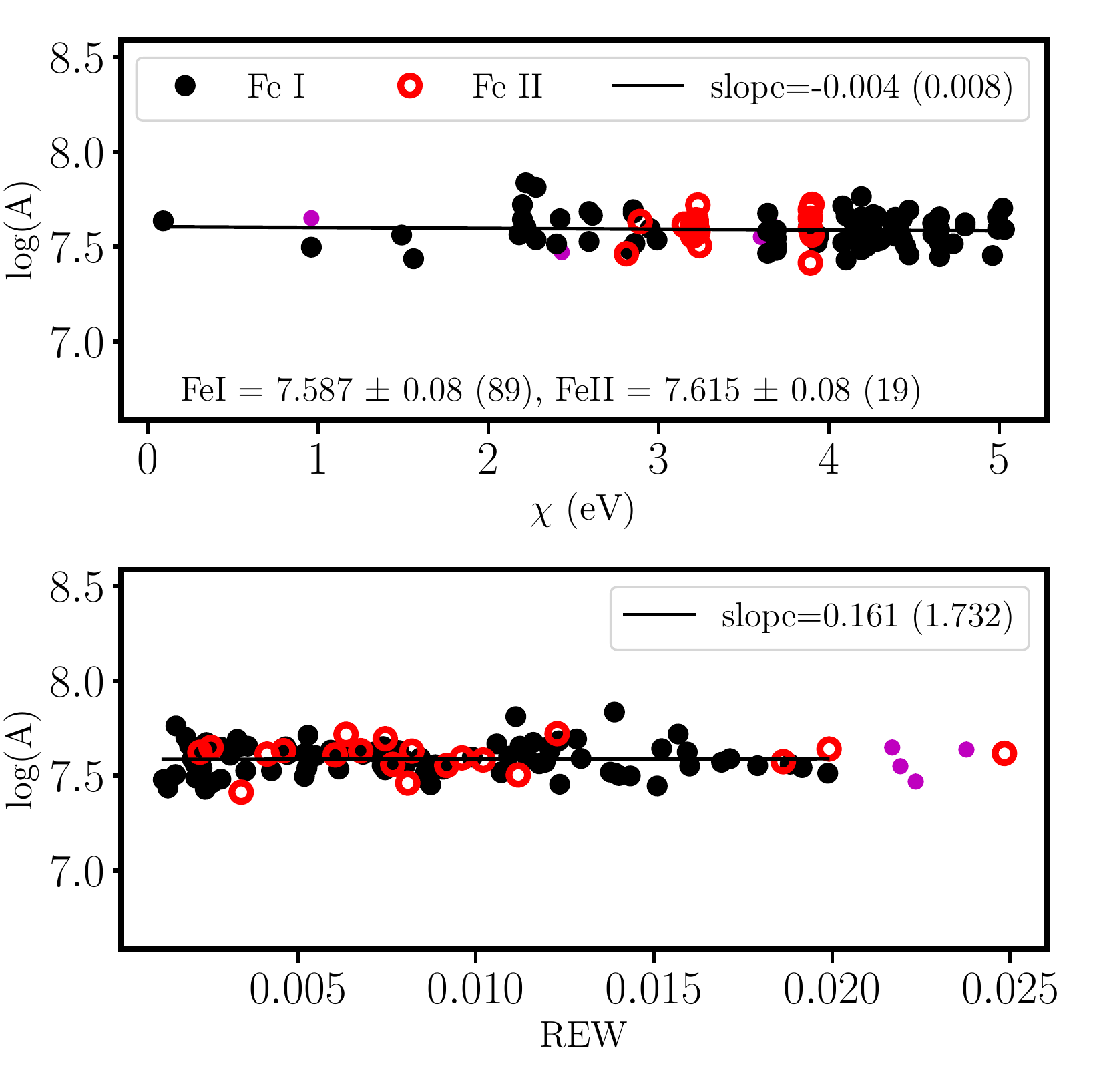}}
\caption{Iron-excitation-ionization balance for KW208 at the adopted atmospheric parameters. The top panel shows the log of the abundance for the Fe\,{\sc i} (black circles) and Fe\,{\sc ii} lines (open red circles) as a function of the excitation potential (in eV) of the line. The bottom panel shows the log abundance for the Fe\,{\sc i} and Fe\,{\sc ii} lines as a function of REW. The magenta circles in both panels represent lines that are stronger than $\approx$120~m\AA.} \label{Fig:FEbalance}
\end{figure}

\begin{table*}[!t]
\center
\caption{Stellar parameters derived from the different techniques}
\begin{tabular}{|l|cccc|cccc|c|}
\hline
 & \multicolumn{4}{c|}{PCA}&\multicolumn{4}{c|}{BACCHUS (adopted)}& {\it Gaia} \\ 
& \Teff & \logg & \met & \vsini & \Teff & \logg & \feh & $\xi_t$ & \Teff \\ 
& (K) & (dex) & (dex) & (\kms) & (K) & (dex) & (dex) & (\kms) & (K) \\ \hline
JS 482 & 4650$\pm$100 & 4.30$\pm$0.15 & 0.11$\pm$0.10 & 7.0$\pm$2.0 & 4658$\pm$110 & 4.55$\pm$0.50 & 0.13$\pm$0.15& 0.69$\pm$0.07& 4685			\\
JS 552 & 	4750$\pm$100 & 4.40$\pm$0.15 & 0.10$\pm$0.10 & 8.0$\pm$2.0 & 4749$\pm$87 & 4.34$\pm$0.22 & 0.12$\pm$0.15 &  1.13$\pm$0.06	&4921	\\
KW23 & 5650$\pm$100 & 4.40$\pm$0.15 & 0.12$\pm$0.10 & 5.5$\pm$2.0 & 5773$\pm$53 & 4.56$\pm$0.24 & 0.20$\pm$0.11 & 1.20$\pm$0.04 & 5601\\
KW30 &5700$\pm$100 & 4.45$\pm$0.15 & 0.10$\pm$0.10 & 6.0$\pm$2.0 & 5716$\pm$45 & 4.57$\pm$0.42 &0.12$\pm$0.12 & 1.18$\pm$0.04& 5597\\
KW208 &5950$\pm$100&4.35$\pm$0.15& 0.12$\pm$0.10 & 9.0$\pm$2.0 & 6005$\pm$19 & 4.46$\pm$0.21 & 0.18$\pm$0.12 & 1.05$\pm$0.04 & 6230 \\
\hline
\end{tabular}\label{results-param}
\end{table*}

The stellar parameters are derived using the standard iron-ionization-excitation balance. More specifically, \Teff\ is derived ensuring that there is no correlation between iron abundance and the excitation potential of the lines. The logarithm of the surface gravity is determined through ionization balance, which amounts to ensuring that there is no significant correlation between Fe\,{\sc i} and Fe\,{\sc ii} abundances. Finally, \vmic\ is determined by requiring that there is no trend  between reduced equivalent width (REW = EW/$\lambda$) and the iron abundances in each line. For more details about BACCHUS and its procedure for stellar parameter derivation, we refer the reader to section~2.2 of \cite{Hawkins2015}. 

We used up to 120 iron lines: up to 90 Fe\,{\sc i} lines and up to 30 Fe\,{\sc ii} lines. To validate the line selection, we derived the parameters of a twilight solar spectrum also taken with ARCES at APO. The solar parameters were found to be 5740 K, 4.40 dex, 0.0 dex and 0.8 km/s for \Teff, \logg, \feh, and \vmic, respectively. 

An illustration of this iron-ionization-excitation balance procedure for one star, KW208, is shown in Figure~\ref{Fig:FEbalance}. In the top panel, we display the logarithm of the abundance of the Fe\,{\sc i} (black circles) and Fe\,{\sc ii} lines (open red circles) as a function of the excitation potential. For KW208, the derived slope in this case is $-$0.004$\pm$0.008. The bottom panel of the same figure displays the logarithm of the abundance as a function of the REW for Fe\,{\sc i} and Fe\,{\sc ii} lines. In this case the slope is 0.161$\pm$1.732.

\subsection{Using Photometry}
In its second data release, {\it Gaia} \citep{gaiamission,gaiadr2} provided data for more than 1.3$\times$10$^9$ objects, including temperatures  based on $G$, $BP$, and $RP$ photometry \citep{gaiadr2phot}. We retrieved the photometric temperatures of our five stars from the {\it Gaia} archive, and they are listed in column 10 of Table~\ref{results-param}. 

We also checked the photometric calibration of \cite{cumming} for the three G stars. \cite{cumming} applied a 10-color combination of the \textit{UBVRI} filters to derive \Teff\ of G stars. Their derived temperatures are 5628 K, 5687 K, and 5979 K for KW23, KW30, and KW208, respectively.\\

Reassuringly, the parameters we derive from these different techniques agree within the error bars. For internal consistency, we adopted the iron-excitation-ionization balance parameters for our stars, as they are based on the same technique that we used for the abundance determinations described below. 

\section{Abundance determination}\label{sec:abun}
\subsection{Approach and Results}
Using the ``abund'' module in BACCHUS, we derived the elemental abundances of up to 24 elements across the light (Li, C, O), $\alpha$ (Mg, Si, Ca, Ti), odd-Z (Na, Al, Sc, V, Ni, Cu, Zn), Fe-peak (Cr, Mn, Fe, Co), and neutron capture (Sr, Y, Zr, Ba, La, Nd) species. Briefly, this module computes the abundances by first fixing the model atmosphere to the one that has a \Teff, \logg, \feh, and \vmic\ matching those derived through iron-excitation-ionization balance. 
It then synthesizes spectra, using \texttt{TURBOSPECTRUM} and accompanying atomic and molecular line lists, with different [X/Fe] abundance ratios for each element independently by up to $\pm$1~dex. 

The abundances for an individual line of a given element are then determined through several methods, including a $\chi^2$ minimization between the observed spectrum and the synthesized spectra and using the EW of the absorption features. All abundances are derived using atomic lines except for carbon, which is a combination of atomic and molecular\footnote{Carbone molecules include CH, C2, and CN.} features. 

To achieve the best possible internal precision, we did a line-by-line differential analysis with respect to the Sun \citep[e.g.,][]{Melendez2009,  Ramirez2009, Hawkins2016}. In this procedure, the abundance for each element and each absorption feature was determined from BACCHUS and then compared directly to the same element and feature in our ARCES twilight solar spectrum. 
This helps to reduce the systematics in the abundance determinations due to incorrect $\log(gf)$ values in the line list, for example. 
\begin{figure*}[!th]
\centerline{\includegraphics[width=2.15\columnwidth]{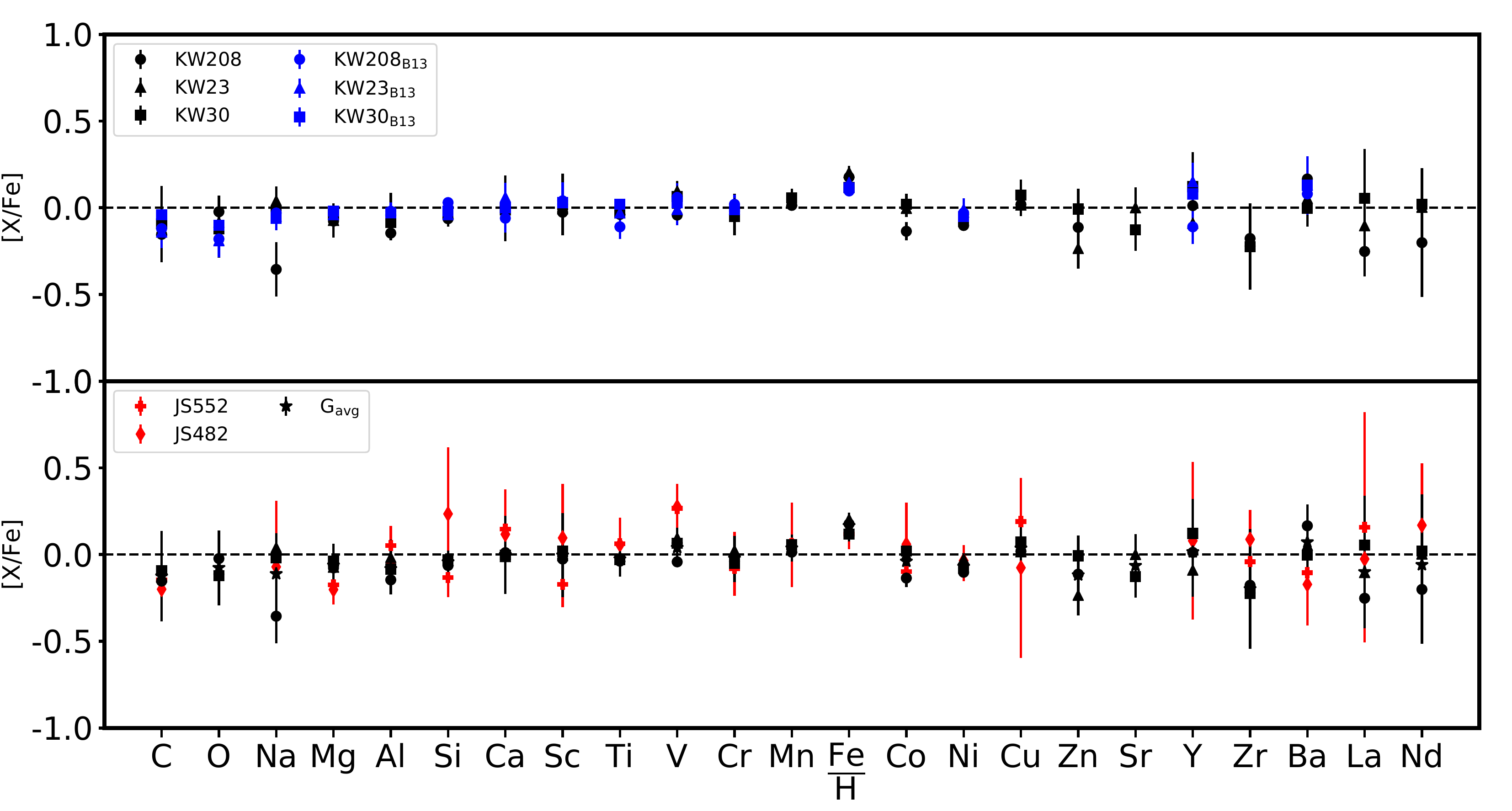}}
\caption{Top: abundance pattern for our three target G stars (black symbols), with abundances derived by \cite{boesgaard2013} for these three G stars shown for comparison (blue symbols). \cite{boesgaard2013} did not derive Mn, Co, Cu, Zn, Sr, Zr, La, or Nd abundances. 
Bottom: abundance pattern for our two target K stars (red symbols), as well as the mean abundances derived for our three G stars (black symbols).}\label{Fig-abunpatternG}
\end{figure*}

The results of this line-by-line differential abundance analysis for the G and K dwarfs are shown in top and bottom panels of Figure~\ref{Fig-abunpatternG}, respectively. For each element, the median of the abundance of all lines is displayed. For comparison, also shown in the top panel of Figure~\ref{Fig-abunpatternG} are the abundances for 15 of these elements determined by \cite{boesgaard2013} for these  three G stars using slightly higher resolution and significantly higher SNR spectra obtained with HIRES on the Keck telescope on Mauna Kea, HI.

\begin{table*}
\begin{center}
\caption{Elemental abundances for our five targets. Columns 7, 8, and 9 give the mean abundances for each element for the three G stars, the two K stars, and their differences, respectively.}
\label{abundance_tab}
\begin{tabular}{l||c|c|c||c|c||c|c|c}
\hline \hline
	&KW23		&KW30		&KW208		& JS 552		&JS 482		&Mean G		&Mean K		&$\Delta$(GK)\\ \hline \hline	
[Fe/H] &	0.20$\pm$0.04   &0.12$\pm$0.04   &0.18$\pm$0.03   & 0.12$\pm$0.03   &0.13$\pm$0.03  &0.17$\pm$0.04 &0.12$\pm$0.01   &0.04$\pm$0.03 \\ \hline
A(Li)  &	2.06$\pm$0.02   &2.12$\pm$0.02   &2.75$\pm$0.02   & \nodata	 	    & \nodata      &2.31$\pm$0.40   & \nodata	         	 &  \nodata	    \\ \hline
[C/Fe] &	-0.13$\pm$0.04   &-0.09$\pm$0.22   &-0.15$\pm$0.13   & -0.15$\pm$0.03   &-0.20$\pm$0.09 &-0.12$\pm$0.03    &-0.17$\pm$0.04   &0.05$\pm$0.03 \\ \hline
[O/Fe] &	-0.08$\pm$0.10   &-0.12$\pm$0.16   &-0.02$\pm$0.09   & \nodata & \nodata  &-0.08$\pm$0.05 & \nodata & \nodata \\ \hline
[Na/Fe]&	0.04$\pm$0.08   &-0.02$\pm$0.05  &-0.35$\pm$0.16    & -0.02$\pm$0.25   &-0.07$\pm$0.38 &-0.11$\pm$0.21   &-0.05$\pm$0.04  &-0.07$\pm$0.05 \\ \hline
[Mg/Fe]&	-0.07$\pm$0.10   &-0.04$\pm$0.05   &-0.08$\pm$0.06  & -0.17$\pm$0.04   &-0.20$\pm$0.09  &-0.06$\pm$0.02  &-0.19$\pm$0.02  &0.13$\pm$0.09 \\ \hline
[Al/Fe]&	-0.02$\pm$0.10  &-0.08$\pm$0.10   &-0.15$\pm$0.02   &0.05$\pm$0.11   &-0.05$\pm$0.12   &-0.08$\pm$0.07 &0.00$\pm$0.07 &-0.09$\pm$0.06 \\ \hline
[Si/Fe]&	-0.03$\pm$0.03   &-0.03$\pm$0.04   &-0.06$\pm$0.05  &-0.13$\pm$0.11   &0.23$\pm$0.38   &-0.04$\pm$0.02 &0.05$\pm$0.26   &-0.09$\pm$0.07 \\ \hline
[Ca/Fe]&	-0.01$\pm$0.08   &0.00$\pm$0.19   &0.01$\pm$0.09   &0.15$\pm$0.10   &0.12$\pm$0.26   & 0.00$\pm$0.01 &0.13$\pm$0.02   &-0.13$\pm$0.09 \\ \hline
[Sc/Fe]&	-0.01$\pm$0.10   &0.02$\pm$0.18   &-0.03$\pm$0.13   &-0.17$\pm$0.13   &0.10$\pm$0.31  &-0.01$\pm$0.02 &-0.04$\pm$0.20   &0.03 $\pm$0.02 \\ \hline
[Ti/Fe]&	-0.01$\pm$0.05   &-0.03$\pm$0.06   &-0.04$\pm$0.07  &0.06$\pm$0.09   &0.06$\pm$0.16   &-0.03$\pm$0.01 &0.06$\pm$0.01   &-0.09$\pm$0.06\\ \hline
[V/Fe] &	0.09$\pm$0.06   &0.07$\pm$0.08   &-0.04$\pm$0.03   &0.27$\pm$0.12  &0.27$\pm$0.13  & 0.04$\pm$0.08  &0.27$\pm$0.01   &-0.23$\pm$0.16\\ \hline
[Cr/Fe]&	0.02$\pm$0.06   &-0.05$\pm$0.11   &-0.04$\pm$0.04  &-0.08$\pm$0.09   &-0.05$\pm$0.18  &-0.02$\pm$0.04 &-0.07$\pm$0.02   &0.05 $\pm$0.03 \\ \hline
[Mn/Fe]&	0.04$\pm$0.05   &0.06$\pm$0.05   &0.01$\pm$0.03    & 0.06$\pm$0.10   &0.06$\pm$0.24   & 0.04$\pm$0.02 &0.06$\pm$0.01   &-0.02$\pm$0.02\\ \hline
[Co/Fe]&	0.00$\pm$0.05  &0.02$\pm$0.06   &-0.13$\pm$0.05    & -0.10$\pm$0.07   &0.06$\pm$0.24  & -0.04$\pm$0.09 &-0.02$\pm$0.11   &-0.02$\pm$0.01\\ \hline
[Ni/Fe]&	-0.03$\pm$0.04   &-0.08$\pm$0.04   &-0.10$\pm$0.02  & -0.09$\pm$0.06   &-0.03$\pm$0.09 &-0.07$\pm$0.04  &-0.06$\pm$0.04   &-0.01$\pm$0.01\\ \hline
[Cu/Fe]&	0.02$\pm$0.06   &0.07$\pm$0.09   &0.02$\pm$0.03    & 0.19$\pm$0.11   &-0.08$\pm$0.52  &0.04$\pm$0.03  &0.06$\pm$0.19   &-0.02$\pm$0.01\\ \hline
[Zn/Fe]&	-0.23$\pm$0.12   &-0.01$\pm$0.12   &-0.02$\pm$0.10  & \nodata 	& \nodata &-0.09$\pm$0.13 & \nodata & \nodata \\ \hline 
[Sr/Fe]&	0.00$\pm$0.12   & -0.13$\pm$0.12   & \nodata       	& \nodata		     & \nodata         	   &-0.06$\pm$0.09 & \nodata  	   	 & \nodata	 	    \\ \hline
[Y/Fe] &	-0.09$\pm$0.12  & 0.12$\pm$0.20   &0.01$\pm$0.11    & \nodata		     &0.08$\pm$0.45   &0.02$\pm$0.11  &0.08$\pm$0.45   &-0.07$\pm$0.05\\ \hline
[Zr/Fe]&	-0.19$\pm$0.13  & -0.22$\pm$0.25   &-0.18$\pm$0.20   &-0.04$\pm$0.14   &0.09$\pm$0.17  &-0.20$\pm$0.02  &0.02$\pm$0.09   &-0.22$\pm$0.16\\ \hline
[Ba/Fe]&	0.05$\pm$0.14   & 0.00$\pm$0.10   &0.17$\pm$0.12    &-0.10$\pm$0.11   &-0.17$\pm$0.24  &0.07$\pm$0.09   &-0.14$\pm$0.13   &0.21$\pm$0.15 \\ \hline
[La/Fe]&	-0.10$\pm$0.07  & 0.05$\pm$0.28 & -0.25$\pm$0.14   & 0.16$\pm$0.66   &-0.02$\pm$0.20  &-0.10$\pm$0.15   &0.07$\pm$0.13   &-0.17$\pm$0.12\\ \hline
[Nd/Fe]&	0.00$\pm$0.15   & 0.02$\pm$0.21 & -0.20$\pm$0.31   & \nodata	& 0.17$\pm$0.36  &-0.06$\pm$0.12 &0.17$\pm$0.36  &-0.23$\pm$0.16\\ \hline
 \hline 
\end{tabular}
\tablecomments{We do not report O, Zn, and Sr abundances for the K stars; see text for details.}
\end{center}
\end{table*}

\begin{table}
\caption{Previous determinations of Praesepe's metallicity}\label{metallicity}
\begin{center}
\begin{tabular}{|c|c|c|}\hline
Authors & Sample & [Fe/H] \\ 
	   &        & (dex) \\
\hline
\cite{1989ApJ...336..798B} & three F stars & 0.09$\pm$0.07\\
\cite{1992ApJ...387..170F} & six F stars & 0.04$\pm$0.06\\
\cite{2007ApJ...655..233A} & four G stars & 0.11$\pm$0.03\\ 
\cite{2008AA...489..403P} & six G stars & 0.27$\pm$0.10\\
\cite{2008AA...483..891F} & six F stars & 0.11$\pm$0.03\\
\cite{2011AA...535A..30C} & three giants & 0.16$\pm$0.05\\
\cite{boesgaard2013} & 11 G stars & 0.12$\pm$0.04\\
\cite{2015AJ....150..158Y} & four giants & 0.16$\pm$0.06 \\
\hline 
\end{tabular}
\end{center}
\end{table}

\begin{table}
\begin{center}
\caption{Comparison of the mean elemental abundances derived from our five targets and those obtained for a sample of 11 Praesepe solar-type stars studied by \citet{boesgaard2013}, which included our three G stars.}
\label{comp_abundance_tab}
\begin{tabular}{l||c|c}
\hline \hline
		& Mean 	& \citet{boesgaard2013} \\ \hline \hline	
[Fe/H] 	& 0.15$\pm$0.08 & 0.12$\pm$0.04 		\\ \hline
A(Li) 	&2.31$\pm$0.40		& \nodata			   	\\ \hline
[C/Fe]	&-0.14$\pm$0.27		& -0.09$\pm$0.05 	\\ \hline
[O/Fe] 	&-0.08$\pm$0.05		& -0.15$\pm$0.09 	\\ \hline
[Na/Fe]	&-0.08$\pm$0.43		& -0.01$\pm$0.03 	\\ \hline
[Mg/Fe]	&-0.11$\pm$0.16		& 0.00$\pm$0.04 	\\ \hline
[Al/Fe]	&-0.05$\pm$0.21		& 0.00$\pm$0.04 		\\ \hline
[Si/Fe] &0.00$\pm$0.40		& 0.00$\pm$0.03 	\\ \hline
[Ca/Fe] &0.05$\pm$0.35		& -0.01$\pm$0.05 	\\ \hline
[Sc/Fe] &-0.02$\pm$0.41		& 0.04$\pm$0.02 		\\ \hline
[Ti/Fe] &0.01$\pm$0.21		& -0.03$\pm$0.06	\\ \hline
[V/Fe]	&0.13$\pm$0.20		& 0.04$\pm$0.04 		\\ \hline
[Cr/Fe] &-0.04$\pm$0.24		& 0.00$\pm$0.03 		\\ \hline
[Mn/Fe] &0.05$\pm$0.27		& \nodata				\\ \hline
[Co/Fe] &-0.03$\pm$0.26		& \nodata				\\ \hline
[Ni/Fe] &-0.07$\pm$0.12		& -0.03$\pm$0.03 	\\ \hline
[Cu/Fe] &0.04$\pm$0.53		& \nodata				\\ \hline
[Zn/Fe] &-0.09$\pm$0.13		& \nodata				\\ \hline 
[Sr/Fe] &-0.06$\pm$0.09		& \nodata				\\ \hline
[Y/Fe] 	&0.03$\pm$0.51		& 0.01$\pm$0.08 		\\ \hline
[Zr/Fe] &-0.11$\pm$0.40		& \nodata				\\ \hline
[Ba/Fe] &-0.01$\pm$0.33		& 0.11$\pm$0.03 		\\ \hline
[La/Fe] &-0.03$\pm$0.75		& \nodata				\\ \hline
[Nd/Fe] &0.00$\pm$0.54		& \nodata				\\ \hline
 \hline 
\end{tabular}
\end{center}
\end{table}

We also tabulated the elemental abundances for each star by taking the median of the abundance for each element over all of the lines where it could be determined. These abundances are reported in Table~\ref{abundance_tab}, as well as the mean abundances for the three G and two K stars and the differences between these mean abundances. 

\subsection{Evaluating the Abundance Uncertainties} \label{subsec:uncert}
To evaluate the uncertainties in our abundances, we first studied the impact of errors in the stellar parameters. For each star and element, we recomputed the abundance while varying \Teff\ by $\pm$100~K, \logg\ by $\pm$0.30~dex, and \vmic\ by $\pm$0.05~\kms. These values were chosen because they represent the typical uncertainty in the stellar parameters for our sample. 

In Table~\ref{abundance_sensitivty_tab}, we indicate the sensitivity of our abundance determinations to the uncertainty in each of the stellar atmospheric parameters. We tabulate the difference in the [X/H] abundance ratio that is caused by a change in the stellar parameters. 

The reported uncertainty for each chemical species is:
\begin{equation}
\sigma_{tot}=\sqrt{\sigma_{[X/H],\Teff}^2 + \sigma_{[X/H],\logg}^2 +\sigma_{[X/H],\xi_t}^2 + \sigma_{\mathrm{mean}}^2 }
\end{equation}
where $\sigma_{\mathrm{mean}}$ is calculated using the classical standard deviation derived from the different abundances of the different lines for each element. $\sigma_{\Teff}$, $\sigma_{\logg}$ and $\sigma_{\xi_t}$ are derived for each element and each star using the sensitivity values of Table~\ref{abundance_sensitivty_tab} scaled to the BACCHUS errors of each fundamental parameter listed in Table~\ref{results-param}. In addition, we report the abundances derived for each absorption feature of every element, along with the $\log(gf)$ of the absorption feature, for each star, in Table~\ref{tab:linebyline}.
 
\section{Discussion and Conclusion}\label{sec:disc}
We have used $R \approx$ 31,000, moderate SNR spectra of five main-sequence G and K stars in Praesepe to test the efficacy of using automated routines to derive the abundances of cooler main-sequence stars and to examine chemical homogeneity in this benchmark open cluster. 

We used two automated, independent approaches to determine the stellar parameters for our five targets: a PCA, which is a statistical spectral fitting method, and the BACCHUS spectral analysis code, which uses iron-excitation-ionization balance. These two approaches produced consistent values for \Teff\ and \logg, which are crucial for robust chemical analysis, despite the relatively low SNR for the K star spectra in particular. For internal consistency, we adopted the BACCHUS-derived parameters for our abundance determinations, also made using BACCHUS.



In Table~\ref{metallicity}, we summarize the findings of a number of spectroscopic studies that derived a metallicity for Praesepe. 
The \cite{boesgaard2013} study, which included the three G dwarfs we observed, found that [Fe/H]~=~0.12$\pm$0.04 for the cluster. This was based on a sample of 11 solar-type cluster members for which these authors obtained $R \approx$ 45,000, SNR~$\approx$~150 spectra. 

Our average derived iron abundance for our three G dwarf targets is 0.17$\pm$0.07 dex. For the two K dwarfs we observed, the average [Fe/H] = 0.12$\pm$0.01. Rather remarkably, averaging the iron abundances we derive for each of our targets gives us [Fe/H] = 0.15$\pm$0.08 for the cluster, consistent with what has been reported in the literature. This is despite our targets having \Teff\ that differ by up to $\approx$1400~K, a wider temperature range than is typical for the stars used in these studies, and despite our spectra having modest SNR compared to those generally obtained by other groups.

To investigate the efficacy of using automated routines to derive the abundances of cooler main-sequence stars, we compared the abundances we derived for the two K stars to those of the G stars. Our abundances agree within $\leq$0.1~dex or less for 13 of the 18 elements that we report for all five stars, providing more evidence that G and K stars in a given open cluster are chemically homogeneous given typical measurement uncertainties. The median difference between the mean G and K stars abundances is 0.08$\pm$0.05 dex, despite serious challenges with the noisier data for the fainter K stars. The quality of the data (and therefore of the BACCHUS measurements)  prevents us from reporting abundances for oxygen, zinc, and strontium. 

As a further test of the robustness of our abundance determination, we compared them to those reported by \cite{boesgaard2013} for the 15 elements these authors report, considering first only our G dwarf targets and then the mean abundances for our five stars. As illustrated by the top panel in Figure~\ref{Fig-abunpatternG}, the agreement between our values for the G stars and those of \cite{boesgaard2013} is generally excellent, with differences of order $\pm$0.05 dex.

In Table~\ref{comp_abundance_tab}, we list the overall mean abundances for our sample and those of \cite{boesgaard2013}. While the uncertainties on our mean values are comparatively large, the overall agreement between these values is very good: the average difference is only about $\pm$0.06~dex.  
In general, the error bars on many of our K star abundances are larger than one would like (reflecting the poor SNR in our data for JS 482 in particular), highlighting the challenges involved in pushing precision abundance determination, whether automated or not, into the cool dwarf regime. 

Still, the overall agreement in the abundances across our sample is encouraging for chemical tagging experiments based on large-scale spectroscopic surveys, which tend to use 4-m class telescopes \citep[e.g., the Galactic Archaeology with HERMES (GALAH) project;][]{2017MNRAS.465.3203M}. It suggests that it may be possible to use automated abundance determination techniques to identify chemically related main-sequence stars across larger temperature ranges than are usually considered in these experiments. Larger samples of higher SNR spectra for cooler stars in benchmark clusters are needed before firmer conclusions can be drawn.



\startlongtable
\begin{deluxetable}{lccc} 
\tablecaption{Abundance determination sensitivity to the stellar parameters. The maximum value of the sensitivity, in dex, is included in this table. \label{abundance_sensitivty_tab}}
\tablecolumns{4}
\tablewidth{0pt}
\tablehead{$\Delta$[X/H] & \Teff$\pm$100~K  & \logg$\pm$0.30~dex  & \vmic$\pm$0.05~\kms } 
\startdata
\multicolumn{4}{c}{KW23}\\
\hline
Li&...&...&...\\
C& 0.06& 0.02& 0.02\\
O& 0.10& 0.09& 0.01\\
Na&0.10&0.07&0.01\\
Mg&0.08&0.11&0.01\\
Al&0.05&0.01&0.01\\
Si&0.01& 0.03&0.01\\
\enddata
\end{deluxetable}

\startlongtable
\begin{deluxetable*}{l|c|c|c|c|c}
\tablewidth{0pt}
\tablecaption{Line-by-line abundances for each element and each of our targets. Column 1 is the star, column 2 is the element, and columns 3, 4, and 5 are the atomic data: line wavelength in~\AA, oscillator strength $\log(gf)$, and excitation potential in eV. The derived absolute elemental abundance is tabulated in column 6.\label{tab:linebyline}}
\tablecolumns{6}
\tablewidth{0pt}
\tablehead{Star	& Element	& $\lambda$ (\AA)	& $\log(gf)$ & EP ($\mathrm{eV}$)	& $\log \epsilon$ } 
\startdata					  
KW23	 &Cr I     &5272.00   &-0.42    &3.449	 &5.430 					  \\ 
KW23	 &Cr I     &5287.17  &-0.87    &3.438	 &5.780 					  \\ 
KW23	 &Cr I     &5296.69  &-1.36    &0.983	 &5.830						  \\ 
KW23	 &Cr I     &5300.74  &-2.0     &0.983	 &5.630 					  \\ 
KW23	 &Cr I     &5304.18  &-0.67    &3.464	 &5.760						  \\ 
KW23	 &Cr I     &5312.85  &-0.55    &3.449	 &5.860 					  \\ 
KW23	 &Cr I     &5318.77  &-0.67    &3.438	 &5.780 					  \\ 
KW23	 &Cr I     &5345.79  &-0.95    &1.004	 &5.780 					  \\
\enddata
\end{deluxetable*}

\acknowledgments
We thank Thomas Masseron for his help. M.A.A.\ acknowledges support provided by the NSF through grant AST-1255419. M.G.\ acknowledges the support of the Fulbright Visiting Scholar Program and the Institute of International Education. M.G.\ thanks the Department of Astronomy at Columbia University for its hospitality. KH is partially supported by a Research Corporation for Science
Advancement TDA grant. 
This work has made use of data from the European Space Agency (ESA) mission \textit{Gaia} (\url{https://www.cosmos.esa.int/gaia}), processed by the \textit{Gaia} Data Processing and Analysis Consortium (DPAC, \url{https://www.cosmos.esa.int/web/gaia/dpac/consortium}). Funding for the DPAC has been provided by national institutions, in particular the institutions participating in the \textit{Gaia} Multilateral Agreement. This work made use of the following software packages: \texttt{astropy} \citep{Astropy2018}, \texttt{astroplan} \citep{Morris2018} and \texttt{aesop} \citep{2018JOSS....3..854M}.

 \software{IRAF \citep{1986SPIE..627..733T}, PHOENIX \citep{Husser2013,Hauschildt3,Hauschildt2,Hauschildt1}, SYNSPEC48 code \citep{spectra}, ATLAS9 code \citep{Kurucz1992}, BACCHUS \citep{Masseron2016}, MARCS \citep{Gustafsson2008}, TURBOSPECTRUM \citep{Alvarez1998, Plez2012}}

\end{document}